\documentclass{svmult}

\usepackage[T1]{fontenc} 
\usepackage[english]{babel}
\usepackage[dvips]{graphics}

\newcommand{\beq}{\begin{equation}}
\newcommand{\eeq}{\end{equation}}
\newcommand{\bma}{\begin{math}}
\newcommand{\ema}{\end{math}}
\newcommand{\beqa}{\begin{eqnarray}}
\newcommand{\eeqa}{\end{eqnarray}}
\newcommand{\ul}\underline

\def\ket#1{|\,#1\,\rangle}

\def\expect#1{\langle\, #1\, \rangle}

\def\opone{\le\textbf{}\textbf{}avevmode\hbox{\small1\kern-3.8pt\normalsize1}}

\begin{document}

\title*{A simple view on the quantum Hall system}

\author{Emil J. Bergholtz and Anders Karlhede}


\institute{Department of Physics, Stockholm University \\ AlbaNova
University Center\\ SE-106 91 Stockholm, Sweden}

\maketitle


\abstract{ The physics of the quantum Hall system becomes very simple when studied on
a thin torus. Remarkably, however, the very rich structure still exists in this limit and there
is a continuous route to the bulk system. Here we review recent progress in
understanding various features of the quantum Hall system in terms of a simple one-dimensional
model corresponding to the thin torus.}



\section{Introduction}

Even though more than twenty years have passed since the
experimental discovery \cite{tsui} of the fractional
quantum Hall effect at filling factor $\nu=1/3$ and its basic explanation due to Laughlin
\cite{Laughlin}, the physics of the quantum Hall regime still
continues to surprise us with new novel phenomena. Already from the
beginning it was clear that the quasiparticles  in the
Laughlin state have fractional charge and later on it was realized that they 
obey fractional statistics \cite{Halp, Aro}. 

Soon after the first observations at $\nu=1/3$ many
other gapped quantum Hall states were observed, some of them at
fractions that could not be explained by Laughlin's wave functions. To
explain these new fractions, hierarchical schemes were developed by Haldane, 
Halperin and Laughlin \cite{hierarchyHald,Halp,hierarchyLaugh} and Jain constructed wave functions for these states 
and proposed an intriguing interpretation
in terms of composite fermions \cite{jain89}, where each of the
electrons captures an even number of magnetic flux quanta,
mapping the original problem of electrons partially filling a Landau
level onto composite fermions filling an integer number of Landau
levels. This gives a nice picture of how the gap responsible for
the quantum Hall effect appears at the fractions $\nu= p/(2mp+1)$ by mapping the system
onto the well understood integer quantum Hall effect. Moreover, the 
composite fermion theory offers an appealing explanation for the existence of
the gapless states observed at even denominator fractions such as $\nu=1/2$,
where the system is mapped onto free fermions in no magnetic field. The mean field 
theory of such states, due to  Halperin, Lee and Read \cite{hlr}, has been spectacularly confirmed by 
surface acoustic wave experiments at $\nu=1/2$ \cite{saw}, and by ballistic experiments near this 
filling factor \cite{ball}. 

However, in our opinion, a microscopic understanding of composite fermions is still lacking 
\cite{dyakonov}.  Gapped quantum Hall states have now been observed that fall outside Jain's main 
scheme \cite{pan}, and the microscopic origin of these
states is under debate. Also, in higher
Landau levels quantum Hall states exist that might possess even
more exotic properties. One such example is the Moore-Read state
\cite{mooreread}, which is believed to describe the quantum Hall
system at $\nu=5/2$ \cite{greiter,morf,rezayihaldane}. This state has attracted
great interest recently due to the supposed non-abelian statistics of
the quasiholes and its possible application to topologically
protected q-bits (decoherence free quantum computational devices)
\cite{freedman}.

In a recent line of research it has been shown that studying the
quantum Hall system on a thin torus allows for both a simple
understanding of already established results and for providing new
insights \cite{bk1, bk2, seidel,marchmeeting,vi,seidel06,natphys}. Here, we give a non-technical
review of this work. References \cite{su1984,Haldane94,bergholtz03} contain relevant 
precursors to the work presented here.

We study the quantum Hall system of spin-polarized electrons on a torus as a function of its
circumference, $L_1$, by mapping the problem onto a one-dimensional
lattice model. When $L_1$ is small, the range of the
electron-electron interaction becomes short (in units of the lattice
spacing), and we get a systematic expansion of the quantum Hall
system around a simple case---the thin torus. The abelian quantum Hall states are
manifested as gapped one-dimensional crystals, 'Tao-Thouless states', and
their fractionally charged excitations appear as domain walls
between degenerate ground states. At half-filling, $\nu=1/2$, the
electrons condense into a Fermi sea of neutral dipoles which 
connects smoothly to the gapless state in the bulk. The non-abelian pfaffian
(Moore-Read) state believed to describe the $\nu=5/2$ phase is
described by six distinct crystalline states, and the non-trivial
quasiparticle and quasihole degeneracies that are crucial for the
non-abelian statistics follow simply from the inequivalent ways of
creating domain walls between these different vacua. This
formulation is manifestly particle-hole symmetric and thus allows
for the construction of both quasiparticles and quasiholes.

The outline of this paper is the following. In section \ref{model}
we set up a one-dimensional lattice model of the lowest Landau
level. In section \ref{thin} we discuss how ground states and
excitations have a very simple and appealing manifestation on the
thin torus, and in section \ref{bulk} we discuss the crucial issue of
how the thin torus picture is connected to the experimentally 
realizable bulk system.

\section{1D lattice model}\label{model}

The energy of a charged particle moving in a magnetic field is
quantized in macroscopically degenerate Landau levels. In the strong
magnetic field limit, the gap between different Landau levels
becomes large and the electrons will populate the lowest available
states. Hence the kinetic energy effectively freezes out, leaving a
strongly interacting problem in the highest partially populated
Landau level (LL). Since a single LL is an effectively
one-dimensional system, it is possible to map the two-dimensional
quantum Hall system onto a one-dimensional problem. It turns out
that this mapping is particularly convenient on the torus.

\begin{figure}[h]
\begin{center}
\resizebox{!}{38mm}{\includegraphics{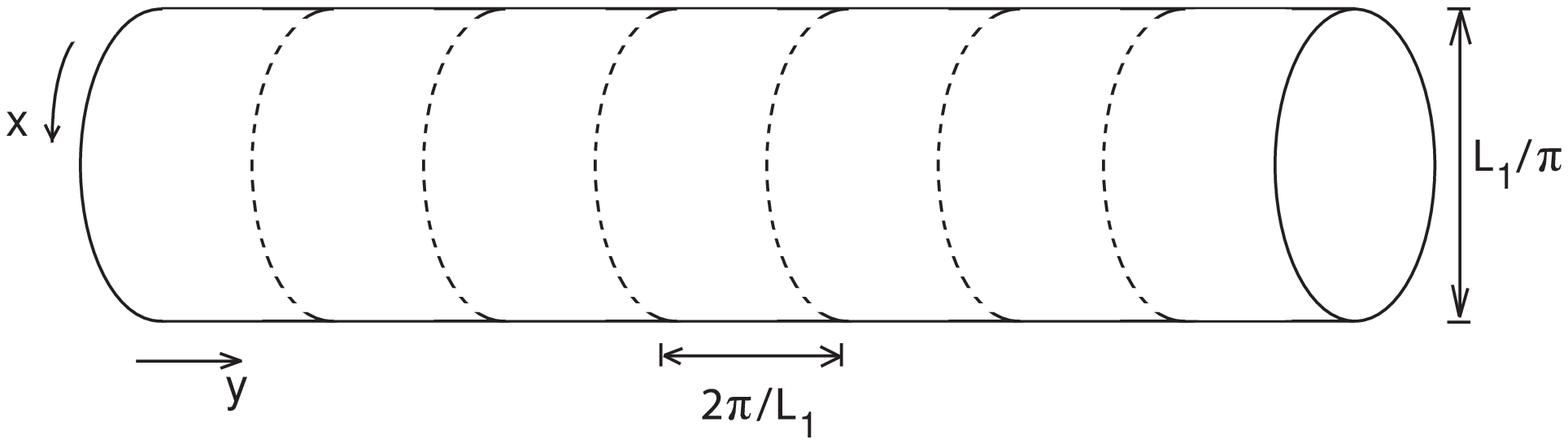}}
\end{center}\caption{ A cylinder with a magnetic field
$\mathbf{B}$ perpendicular to its surface. The single particle
states are centered along the lines $y_m=-2\pi m/L_1$ and
can be thought of as sites in a one-dimensional lattice.}
\label{cylinder}
\end{figure}

For simplicity we consider the problem of an electron moving in a
perpendicular magnetic field on the surface of a cylinder (the torus
case is obtained by straight forward periodising). In Landau gauge,
$\mathbf A=  By\hat x$, the lowest Landau level states are
\begin{equation}
\psi_{m}(\mathbf{r})=\frac{1}{\pi^{1/4}L^{1/2}}e^{2\pi
imx/L_1}e^{-(y+2\pi m/L_1)^{2}/2},\label{cylstate}\end{equation}
where we use units such that $\ell = \sqrt{\hbar c/eB}=1, \hbar=1$,
and label the states by integers $m$. The states are centered along
the lines $y_m=-2\pi m/L_1$, given by the momentum in the
$x-$direction. This provides an explicit mapping of the
two-dimensional electron gas in the lowest Landau level onto a
one-dimensional lattice model, where the lines $y_m$ can
be thought of as the sites, see Figure \ref{cylinder}.\\\\

A general (two-body) interaction Hamiltonian takes the form \beqa
\label{ham} H=\sum_n \sum_{k > m}
V_{km}c^\dagger_{n+m}c^\dagger_{n+k}c_{n+m+k}c_n ,\label{hint} \eeqa
where $V_{km}$ are matrix elements that can be calculated for a
given real-space interaction. The physics of the interaction can be
understood by dividing $H$ into two parts: $V_{k0}$, the
electrostatic repulsion (including exchange) between two electrons
separated $k$ lattice constants, and $V_{km}$, the amplitude for
two particles separated a distance $k-m$ to hop symmetrically to a
separation $k+m$ and vice versa. The symmetry of the hopping, which is a 
consequence of conservation of momentum, implies 
that the position of the  center of mass is conserved. 

A general $N_e-$particle state in the lowest Landau level is a
linear combination of states characterized by the positions (or,
equivalently, the momenta) at which they are centered. We
represent these (Slater determinant) states in Fock space as
$\ket{n_1n_2n_3\ldots}$ where $n_i=0,1$ according to whether site
$i$ is occupied or not. The problem of finding the
ground state and the low lying excitations, at filling fraction
$\nu=N_e/N_s$, is thus a matter of
arranging $N_e$ electrons on $N_s$ sites.

A very important property of the obtained lattice model is that the
lattice constant is $2\pi/L_1$. This means that, for a given
real space interaction, the interaction in the one-dimensional lattice model becomes short range
in units of the lattice spacing when the torus becomes thin and we can hope to be able
to solve the problem in this limit. The experimental situation, on
the other hand, is obtained as $L_1\rightarrow \infty$, where the
lattice model becomes infinitely long range measured in units of the lattice
constant. When the system is studied as a function of $L_1$, we find
that many of the characteristic features of the quantum Hall system
is independent of $L_1$ and there is a continuous route between the
two extreme cases---we claim that the two cases are adiabatically connected.

\section{The thin torus}\label{thin}

Here we consider the quantum Hall system at generic filling
fractions, $\nu=p/q<1$, in the limit $L_1\rightarrow 0$. For reasonable
interactions (including Coulomb), the problem becomes a classical
electrostatic one-dimensional problem and the ground states are
regular lattices of electrons where the particles are as far apart as possible, as
shown in Table \ref{gstable}.
\\

\begin{table}[h]

\begin{center}
\begin{tabular}{ l }
\hskip 19 pt $\ket{\ul{100}100100100100100\dots}$ \ \ \ $\nu=1/3$\\
\hskip 19 pt $\ket{\ul{10100}101001010010100\dots}$\ \ \ $\nu=2/5$\\
\hskip 19 pt $\ket{\ul{10100100100}10100100100\dots}$\ \ \ $\nu=4/11$\\
\end{tabular}
\end{center}
\caption{Examples of ground states in the thin limit,
$L_1\rightarrow 0$. The underlined unit cells containing $p$
electrons on $q$ sites are periodically repeated in the $\nu=p/q$
ground state. The $q$-fold degeneracy on the torus is reflected by
$q$ different translations of the unit cell.} \label{gstable}
\end{table}

The reason that the physics is completely determined by electrostatics in the thin limit
is actually rather simple. The single particle states are essentially
gaussians extended roughly one unit length ({\it i.e.} one magnetic
length) and separated by the lattice constant $2\pi/L_1$.
Consequently, the overlap between different one-particle wave functions becomes very
small and the only non-vanishing matrix elements are those where each electron is created 
and destroyed at the same site, {\it i.e.} the electrostatic matrix elements $V_{k0}$. 
Thus, $L_1$ is a parameter that controls the strength of the hopping, which can 
be continuously turned on by increasing $L_1$.

The ground states in the thin limit are regular lattices with unit
cells containing $p$ electrons and $q$ sites at filling $\nu=p/q$.
This is true for any repulsive interaction that is monotonic, with
positive second derivative---Coulomb falls
into this category. The same ground states were obtained by Hubbard when
he investigated generalized Wigner lattices in the seemingly very
different context of quasi-one-dimensional salts \cite{hubbard}. It
is interesting to note that, at $\nu=1/3$, the thin limit ground state, 
see Table \ref{gstable}, is the state
originally proposed by Tao and Thouless in 1983 to explain the
fractional quantum Hall effect \cite{tt}. We call these states, at 
general filling factor, Tao-Thouless (TT) states, stressing the fact that they are 
different from 
ordinary, classical crystals or Wigner crystals.
It is important to note that the TT-states have a gap to all
excitations---there are no phonons. The reason for this is that
once the fluxes through the holes of the torus are fixed, then the
positions of the one-particle states along the torus are fixed, and
hence no vibrations of the lattice are possible. Note also that the
$q-$fold degeneracy, present for all energy eigenstates on the torus
\cite{haldane}, is trivially manifested by the $q$ different
translations of the unit cell.

\subsection{Gapped fractions and fractional charge}\label{qp}

At odd denominator fractions in the lowest Landau level, the 
TT-states describe (but are extreme forms of) the gapped  abelian quantum Hall
states observed in the laboratory. In section \ref{bulk} we discuss
this connection further, but let us first consider the structure of
ground states and fractional charge that emerge in the thin limit.

At the Jain fractions, $\nu=p/(2pm+1)$, the unit cells are
$10_{2m}(10_{2m-1})_{p-1}$ in chemical notation. At
$\nu=1/3$ the unit cell is $100$, at $\nu=2/9$ it becomes
$100001000$ and so on. These states are
gapped and $q-$fold degenerate.

The low energy excitations of the TT-states at arbitrary filling fractions are
domain walls separating sequences of degenerate ground states. These domain
walls carry fractional charge and correspond to the quasiparticle and quasihole
excitations in the bulk.

At $\nu= 1/q$ a quasihole (quasiparticle) is constructed by
inserting (removing) a zero somewhere in the ground state, see Table
\ref{thirdcharge}. This is very similar to Laughlin's original concept
of creating a quasihole by inserting a flux quantum. At  $\nu=
p/(2pm+1)$ the corresponding quasiparticle (quasihole) excitations
are obtained by inserting (removing) $10_{2m-1}$ somewhere in the
TT-state with unit cell $10_{2m}(10_{2m-1})_{p-1}$.\\

\begin{table}[htdp]

\begin{center}
\begin{tabular}{ l }
\hskip 19 pt $\ket{100100100100100100100100100100100\dots}$\\
\hskip 19 pt $\ket{100\ul{101}00100100\ul{101}00100100\ul{101}00100\dots}$\\
\hskip 19 pt $\ket{1001\ul{000}1001001\ul{000}1001001\ul{000}100100\dots}$\\
\end{tabular}
\end{center}
\caption{The $\nu=1/3$ ground state, and the corresponding states
with three quasiparticles and three quasiholes respectively. Note
that the underlined concentration of electrons (or holes) are domain
walls between degenerate $\nu=1/3$ ground states. The charge ($\pm e/3$) of these
excitations is determined by Su and Schrieffer's counting argument.} \label{thirdcharge}
\end{table}

\begin{table}[htdp]

\begin{center}
\begin{tabular}{ l }
\hskip 19 pt $\ket{1010100101010010101001010100\dots}$\\
\hskip 19 pt $\ket{1010100\ul{101010}10010101001010\ldots}$\\
\hskip 19 pt $\ket{1010100\ul{10}1001010100101010010\ldots}$\\
\end{tabular}
\end{center}
\caption{The $\nu=3/7$ ground state, and the corresponding states
with a quasiparticle and a quasihole respectively. Note that
inserting/removing $10$ creates domain walls with the correct charge
$\pm e/7$. (Inserting/removing $100$ would instead create domain
walls with charge $\pm 2e/7$.)} \label{37}
\end{table}

The charge of these excitations is determined by Su and Schrieffer's
counting argument \cite{counting}. By removing $10_{2m-1}$ at
$2pm+1$ separated position and adding $2m$ unit cells
$10_{2m}(10_{2m-1})_{p-1}$ to keep the number of sites fixed,
$2pm+1$ quasiholes, each with charge
$e^*=e\frac{(2pm+1)-2pm}{2pm+1}=\frac{e}{2pm+1}$, are created. This
readily generalizes to generic fillings $p/q$, where
the lowest lying excitations naturally emerge as domain walls
carrying charge
\begin{equation}e^*=\pm\frac{e}{q}.\end{equation}

\subsection{The non-abelian pfaffian state}

The single particle states differ from (\ref{cylstate}) in the higher Landau levels,
thus the interaction ({\it i.e.} $V_{km}$) is different, and as a
consequence, the ground states and their excitations may differ from
those in the lowest Landau level. Perhaps most notably, the ground
state at half-filling in the second Landau level appears to be gapped and is believed to be accurately
described by the Moore-Read pfaffian state \cite{mooreread}. This state,
which is motivated by conformal field theory, has quasihole
excitations with charge $e/4$ that can only be created in pairs, and
obey non-abelian statistics.

Here we describe how this state is manifested on the thin torus and
give the degeneracies of the quasihole excitations that are
crucial for the existence of non-abelian statistics. Moreover, the
particle-hole symmetry allows us to construct also 
quasiparticles, as well as states with general combinations of quasiholes and
quasiparticles \cite{vi}.

The pfaffian states on the torus are known to be the exact ground
states of a hyper-local three-body potential \cite{greiter,
rezayihaldane}. In the thin torus limit, this implies that the
electrostatic energy (of this three-body potential) is minimized by
separating all triples of particles as much as possible. At
half-filling this means that there are no sequences of four
consecutive sites containing three electrons (or holes). The six
states displayed in Table \ref{pfaffvac} are the unique states
at half-filling that have no such sequences.

\begin{table}[htdp]
\begin{center}
\begin{tabular}{ l }

 $\ket{010101010101\dots}$  \ \   2 translations \\
 $\ket{001100110011\dots}$  \ \  4 translations

\end{tabular}
\end{center}
\caption{The six degenerate pfaffian ground states on a thin torus.}
\label{pfaffvac}
\end{table}
The extra freedom created by the additional pfaffian ground states
allows for the creation of domain walls carrying charge $e^*=\pm
e/4$---{\it i.e.} half of the fractional charge $e^*=\pm e/2$ that
is implied by the center of mass degeneracy. The domain walls that achieve this are
those between the two different kinds of ground states
$\ket{10101010\ldots}$ and $\ket{11001100\ldots}$, as shown in Table
\ref{halfholes}. Again this charge is readily
determined by Su and Schrieffer's counting argument. Note also
that, because of the periodic boundary conditions, these excitations 
can only be created in pairs.\\

\begin{table}[htdp]

\begin{center}
\begin{tabular}{ l }
\hskip 19 pt $\ket{0101\ul{0100}110011\ul{0010}10101\dots}$ \ \ \ two quasiholes  \\
\hskip 19 pt $\ket{01010\ul{1011}001100\ul{1101}0101\dots}$ \ \ \ two quasiparticles   \\
\hskip 19 pt $\ket{01010\ul{1011}0011\ul{0010}10101\dots}$ \ \ \  a quasiparticle-hole pair
\end{tabular}
\end{center}
\caption{Examples of domain walls with fractional charge $\pm e/4$.}
\label{halfholes}
\end{table}

The degeneracy of these excitations is readily determined by 
considering the various ways of matching the domains. In Ref. 
\cite{vi} we derived that the degeneracy of a state with $2n-k$
quasiholes and $k$ quasiparticles with fixed positions is $2^{n-1}$. 
Results similar to ours have also been obtained by Haldane \cite{marchmeeting},
and subsequently also by Seidel and Lee \cite{seidel06} for the
closely related bosonic pfaffian state at $\nu=1$.

\subsection{The half-filled Landau level}\label{half}

The physics of the half-filled lowest Landau level is known to be
very different from the gapped fractions discussed above.
There is strong experimental and numerical evidence that the
system is gapless. In the composite fermion picture, all magnetic flux is
attached to the electrons and the system becomes a free Fermi gas
of composite fermions in no magnetic field \cite{jain89,hlr,RR}. Furthermore,
it has been proposed that the quasiparticles are dipoles \cite{dhlee,
read98, pasq}.

In the thin limit, the $\nu=1/2$ ground state is
$|1010101010....\rangle$ and the (gapped) low lying excitations are
the fractionally charged excitations described above. In fact, the
$\nu=1/2$ state has a larger energy gap than the $\nu=1/3$ state on
the thin torus. This is clearly different from the observed gapless
state in the bulk.

In order to explain this discrepancy we consider the situation when
$L_1$ increases from zero. Short range hopping terms will now become
important and start to compete with the electrostatic terms.
However, the shortest range hopping $V_{21}$ annihilates the TT
state $|1010101010....\rangle$. Also, from early numerical
investigations it was clear that there is a sharp transition from
the TT-state $|1010101010....\rangle$ at $L_1\sim 5.3$ to a gapless homogeneous 
state \cite{bergholtz03}. 

It is interesting to contrast $\nu=1/2$ with $\nu=1/3$. At  $\nu=1/2$, the ground state is the
TT-state $\ket{101010\ldots}$ when $L_1 \rightarrow 0$. As noted, this state is  annihilated by the shortest range hopping term 
$V_{21}$ which favours hoppable states of the type $\ket{11001100\ldots}$ Thus there is a competition between the 
electrostatic terms and the hopping term and this leads to a phase transition to a gapless state when $L_1$ grows. 
For $\nu=1/3$ on the other hand, the TT-state $\ket{100100\ldots}$ favoured by electrostatics is also a maximally hoppable state 
favoured by the short range hopping term. In this case there is no competition between electrostatics and hopping and
there is no phase transition as $L_1$ grows. 

\begin{figure}
\begin{center}
\resizebox{!}{90mm}{\includegraphics{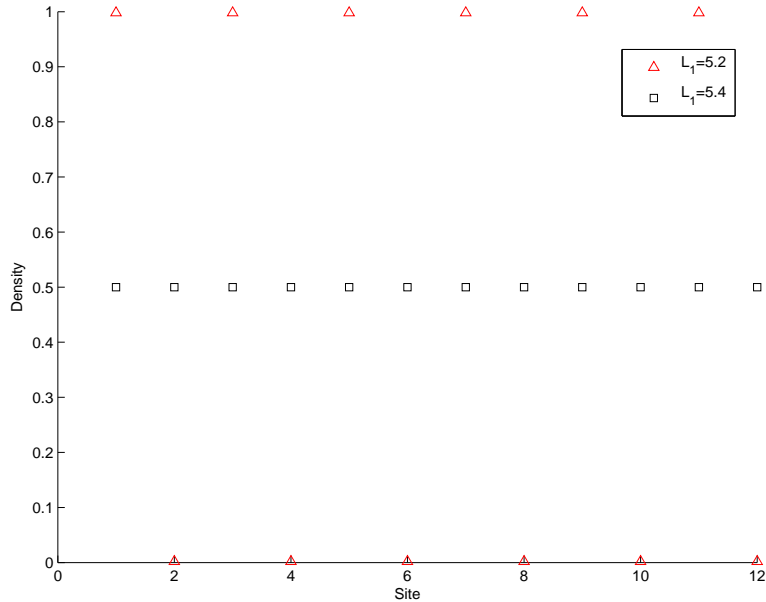}}
\end{center}\caption{The evolution of the one-dimensional density 
$\langle c^\dagger _k c_k \rangle $ from the small
      $L_1$ TT-state (triangles) to the homogenous state
      (squares) at $\nu=1/2$. At $L_1\sim 5.3$ there is a sharp
transition from the TT-state to a homogenous state that is described
by our solvable model, and corresponds to a Luttinger liquid of
neutral dipoles. At the transition the quantum numbers change.} \label{dens6_12}
\end{figure}

We now briefly discuss a solvable model that accurately
describes the system at $L_1$ slightly larger than $5.3$.
The low-energy sector of the model consists of free one-dimensional neutral
fermions (dipoles) \cite{bk1}. The crucial part in the hamiltonian
turns out to be the hopping term $V_{21}$---the other terms can be
treated as perturbations yielding an interacting Luttinger liquid.

We start with the hamiltonian \beqa \label{hamtrunc} H^*= -V_{21}
\sum_{n} c^\dagger_{n}c_{n+1}c_{n+2}c^\dagger_{n+3} + H.c. \ \ .
\eeqa This provides a good approximation of the interaction on a
thin, but not infinitely thin, torus ($L_1\sim 6$) as discussed in Ref. \cite{bk1}.

We define a subspace ${\cal H}^\prime$ of the full Hilbert space by
requiring each pair of sites $(2p-1,2p)$ to have charge one (the
equivalent  grouping of the sites  $(2p,2p+1)$ gives a trivial copy
of our solution). In Ref. \cite{bk1} it is argued that ${\cal
H}^\prime$ contains the low-energy sector under fairly general
conditions. It agrees with what we find in numerical studies, and
${\cal H}^\prime$ contains the maximally hoppable state
$|100110011001....\rangle$. Furthermore, $H^*$ preserves the
subspace $\cal H^\prime$, thus any other ground state candidate may
not mix with the states in $\cal H^\prime$.

There are two possible states for a pair of sites in $\cal H
^\prime$;
\begin{equation}|\downarrow \rangle \equiv |01\rangle, \ \ |\uparrow \rangle
\equiv |10\rangle\end{equation} and it is natural to introduce the
spin operators \begin{equation}s^+_p=c^\dagger_{2p-1}c_{2p},\ \ \
s^-_p=c^\dagger_{2p}c_{2p-1}\label{dipole}.\end{equation} On states in $\cal H
^\prime$, $s^+, \, s^-$ describe hard core bosons---they commute on
different sites but obey anti-commutation relations on the same
site. In this subspace, $H^*$ is simply the nearest neighbor spin 1/2 $XY$-chain,

\begin{equation} H^*=V_{21}\sum_{p} (s^{+}_{p+1} s^{-}_{p}+s^{-}_{p+1}
s^{+}_{p}).\end{equation} The (hard core) bosons can be expressed in
terms of fermions $d$ using the Jordan-Wigner transformation,
\begin{equation}s^-_p=K_p d_p,\ \ \ K_p=e^{i\pi
\sum_{j=1}^{p-1}d^\dagger_j d_j},\end{equation} and the Hamiltonian
(\ref{hamtrunc}) is then that of free fermions.

The ground state is obtained by filling all the negative energy
states. The excitations are neutral particle-hole excitations out of
this Fermi sea. These excitations have a natural interpretation in
terms of dipoles as is seen from (\ref{dipole}), and in the limit $N_e\rightarrow \infty$, the
excitations become gapless. It is also straight forward to show that
the state is homogeneous. We would like to stress that this explicitly and exactly maps 
(the low energy sector of) a system of strongly interacting electrons in a strong magnetic field onto 
a system of non-interacting particles that are neutral and hence are unaffected by the magnetic field. 

By considering the relation between the real system---where the
electrons interact via Coulomb repulsion---and our model, we
conclude that the $\nu=1/2$ system is a Luttinger liquid of these
dipoles on a thin torus ($L_1$ slightly larger than $5.3$). This conclusion is
supported by numerical calculations for both Coulomb
\cite{bk2} and short range interactions \cite{bergholtz03}. Note
also that the obtained solution has striking similarities to the
bulk state---both are homogenous gapless states with quasiparticles
(dipoles) that do not couple to the magnetic field.

\section{Bulk physics}\label{bulk}
In this section we discuss how the two-dimensional bulk physics is related to the
physics in the thin limit. We will argue that the abelian and
non-abelian gapped states, as well as the gapless state at $\nu=1/2$,
are adiabatically connected to the states found on the thin torus. The strength of 
the argument varies with the filling factor but we believe the over all picture of bulk states at 
generic filling factor being adiabatically connected to simple ground states on the thin torus 
is firmly established. 

Before we proceed with a more detailed account for each of the considered cases we make 
two important remarks: 1) The TT-states and the bulk QH-states do in fact have the the same symmetries and qualitative properties. 
That the TT-state is not homogenous is not a result of spontaneous symmetry breaking---in fact the Laughlin/Jain states have periodic density variations on any finite torus \cite{haldanewfs}. 2) As indicated in section \ref{half} , there is actually a simple way of understanding why the TT-state melts at half-filling while it develops 
smoothly into the bulk QH-state at {\it e.g.} $\nu=1/3$.

\subsection{Abelian states}
We begin by considering the simplest case,
$\nu=1/q$, $q$ odd. At these filling factors the Laughlin wave functions 
describe the bulk physics; moreover, they are the exact and unique ground states to a short range pseudo-potential 
interaction and there is a gap to all excitations \cite{pseudo, Kivelson}. This holds also 
on a torus (or cylinder) for arbitrary circumference $L_1$\footnote{On the torus, the ground state of course has the trivial 
$q$-fold center of mass degeneracy.}. This is fairly obvious since it depends only on 
the short distance property of the electron-electron interaction. In our opinion, this 
establishes that the ground state develops continuously as $L_1$ increases, without a phase transition, 
from the  TT-state  to the bulk Laughlin state for this short range interaction. This result is implicit in the 
work of Haldane and Rezayi \cite{Haldane94}. The same is then very likely to be true for the 
Coulomb interaction---this is supported by exact diagonalization where no transition is seen as $L_1$ varies. 

We now show that the Laughlin wave function on a cylinder 
\begin{equation}
\Psi_{1/q}=\prod_{n<m}(e^{2\pi i z_n/L_1}-e^{2\pi i z_m/L_1})^q
e^{-\frac 1 2 \sum_n y^2_n}, 
\end{equation} 
where $z=x+iy$, approaches the TT-state as the radius of 
the cylinder shrinks \cite{Haldane94}. 
Expanding $\Psi_{1/q}$
in powers of $e^{2\pi i z/L_1}$ and using that the single particle
states (\ref{cylstate}) can be written as
$\psi_k=\frac{1}{\pi^{1/4}L_1^{1/2}}(e^{2\pi i z/L})^k
e^{-y^2/2}e^{-2\pi^2k^2/L_1^2}$, one finds
\begin{eqnarray} \Psi_{1/q}=\sum_{\{k_n\}}\prod_n
c_{\{k_n\}}(e^{2\pi i z_n/L_1})^{k_n}e^{-\frac 1 2 \sum_n
y^2_n}=\nonumber\\=\frac{1}{\pi^{N_e/4}L_1^{N_e/2}}\sum_{\{k_n\}}
\!c_{\{k_n\}}\!\psi_{k_1}\psi_{k_2}\!\cdots\psi_{k_{N_e}} e^{2 \pi^2
\sum_n k^2_n/L_1^2},\end{eqnarray} 
where $c_{\{k_n\}}$ are
coefficients that are independent of $L_1$. The weight of
a particular electron configuration is multiplied by the factor
$e^{2 \pi^2 \sum_m k^2_m/L_1^2}$, thus in the limit $L_1\rightarrow
0$ the term with the maximal $\sum_m k^2_m$ will dominate (all terms have the same $\sum_m k_m$). The
dominant term is the one that corresponds to the TT-state
discussed above, where the electrons are situated as far apart as
possible. In this case at every $q$:th site. This argument can be generalized to the Jain wave functions 
describing the ground states at filling factors $\nu=p/(2mp+1)$ showing that they approach the TT-states above as $L_1
\rightarrow 0$. It can also be generalized to show that  the  fractionally charged quasiparticles  in 
the TT-state, discussed in Section \ref{qp},  are the $L_1 \rightarrow 0$ limits of the bulk quasiparticles at filling factor $\nu=p/(2mp+1)$. 

\begin{figure}[h]
\begin{center}
\resizebox{!}{90mm}{\includegraphics{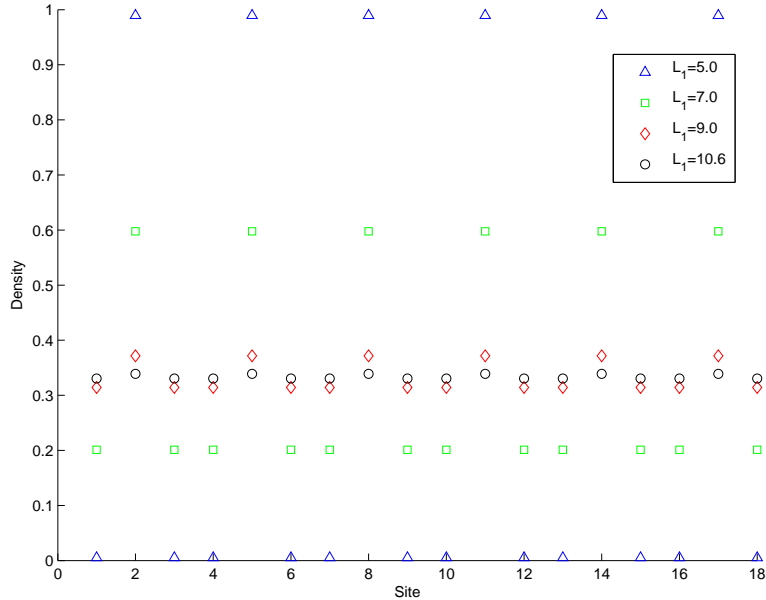}}
\end{center}\caption{The evolution of the one-dimensional density  
$\expect{c^{\dag}_{n}c_{n}}$ from the small
      $L_1$ TT-state (triangles) to the nearly homogenous bulk Laughlin state
      (circles) at $\nu=1/3$. This process is smooth and the quantum
      numbers  $K_\alpha$ remain unchanged as $L_1$ changes.
      Results are obtained from exact diagonalization of an
      unscreened Coulomb potential. }
\label{dens6_18}
\end{figure}

The TT-state and the bulk Laughlin/Jain state on the torus at $\nu=p/(2mp+1)$ have 
the same quantum numbers. The symmetry generators that
commute with the hamiltonian are $T_1$ and $T_2$ ($T_\alpha$
translates all particles in the $\alpha$-direction). The Laughlin/Jain
state is an eigenstate of $T_1$ and $T_2^{2mp+1}$, with quantum numbers $K_1$ and $K_2$, whereas $T_2^k,
k=1,2,\ldots,2mp$ generate the degenerate states---this is true for any 
$L_1$---and the eigenvalues are independent of $L_1$.
The state is inhomogeneous for any finite $L_1$, although the
inhomogeneity decreases very rapidly as $L_1$ grows. Furthermore,
the TT-state and the Laughlin/Jain state both have a gap and have
quasiparticles and quasiholes with the same fractional charge. 

The conclusion is that there is no phase transition separating the 
TT-states and the bulk Laughlin states. This result has a long history.  The very first observation was
made already in 1984 by Su who discussed the TT-state as the 'parent
state' of the Laughlin state and observed that the fractionally charged quasiparticles 
could be thought of as domain walls between the degenerate vacua. Rezayi and Haldane
noted that the Laughlin state is the exact ground state for the short range interaction on a cylinder of any 
circumference and showed that the state
approaches a crystal as $L_1\rightarrow 0$ in 1994 \cite{Haldane94}.
More recently this was reexamined by the present authors in DMRG calculations
\cite{bergholtz03,bk1} and in exact diagonalization \cite{bk2} and a careful numerical study of the
rapid crossover from the  TT- state to a virtually homogeneous state was performed by 
Seidel {\it et. al.} using Monte Carlo methods \cite{seidel}.

In the case of the Jain states, there is no known interaction which
they are the exact and unique ground states of. However, as we have noted 
above they have the same qualitative properties as the corresponding TT-states: same
quantum numbers, gap and quasiparticles with the same charge.  These TT-states, including quasiparticle 
excitations, are obtained 
as the $L_1 \rightarrow 0$ limits of Jain's wave functions. 
Furthermore,  exact diagonalization of small systems show a smooth development 
of the ground state from the TT-state to the Jain state as $L_1$ grows. No 
transition  is observed and there is a gap for all $L_1$
\cite{bk2}. Recent progress strongly suggests that this picture is true also for
more general odd denominator fractions in the lowest Landau level,
such as the non-Jain state at $\nu=4/11$; in these cases no phase transition is observed for small systems
and a new set of trial wave functions  connect the solvable limit to the bulk \cite{natphys,hans}.  
We conclude that the adiabatic continuity holds also for 
the hierarchy states.

\subsection{Non-abelian states}

Recently, it has been understood that also non-abelian gapped
quantum Hall states follow the same pattern as we outlined for
the abelian states above \cite{marchmeeting, vi, seidel06}. 

The six Moore-Read pfaffian ground states \footnote{There are three distinct pfaffian
wave functions on the torus. This together with the two-fold center
of mass
degeneracy gives all the six states on the thin torus.} are the
exact ground states of a hyperlocal three-body interaction on the
torus---as in the case of the Laughlin states, this holds for
general $L_1$ as it depends on the local properties only. As
$L_1$ decreases the states continuously approach the TT-states in
Table \ref{pfaffvac}.

\subsection{The gapless state at $\nu=1/2$}

The $\nu=1/2$ solution on the thin torus, discussed above, has striking
similarities to what is expected from theory and experiment for the
bulk state. Based on this, we conjectured \cite{bk1} that this state
develops continuously, without a phase transition, to the bulk state
as $L_1 \rightarrow \infty$. This is however a much more delicate issue than
it is for the states above since the state at $\nu=1/2$ is gapless.

To investigate this conjecture, we performed exact diagonalization
studies of small system for various $N_e$ and $L_1$ using an
unscreened Coulomb potential \cite{bk2}. The obtained ground states
were then compared with the Rezayi-Read state \cite{RR}, that is
expected to describe the bulk state, by calculating overlaps. 
On the torus the Rezayi-Read wave function
takes the form
\begin{eqnarray}
\Psi_{RR}  = {\rm det_{ij}}[e^{i{\bf k}_i\cdot {\bf R}_j}] \Psi_{\frac 1 2}  \ \ \ ,
\end{eqnarray}
where ${\bf R}$ is the guiding center coordinates and $\Psi_{\frac 1 2}$ is the
bosonic Laughlin state at $\nu=1/2$. This wave function depends on a set
of momenta $\{{\bf  k}_i \}$, which determine the conserved quantum numbers $K_\alpha$.

\begin{figure}[h!]
\begin{center}
\resizebox{!}{30mm}{\includegraphics{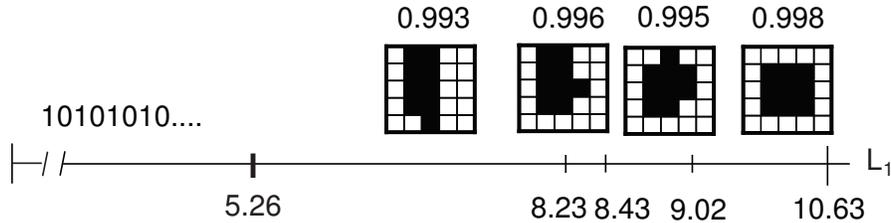}}
\end{center}\caption{'Phase diagram' showing the ground states for $\nu=1/2$
      as a function of $L_1$ for nine electrons \cite{bk2}. The results are obtained in exact
diagonalization, using unscreened Coulomb
      interaction. Overlaps with the Rezayi-Read state with the
      displayed Fermi seas of momenta are shown above each Fermi sea.}
\label{phasediagrams}
\end{figure}

For $L_1\leq 5.3$ the ground state is the TT-state
$\ket{10101010\ldots}$. At $L_1\sim 5.3$ there is a sharp transition
into a new state that we identify as our Luttinger liquid solution,
discussed above. As $L_1$ is increased further, there is a number of
different transitions to new states, but these transitions are all
much smoother than the one at $L_1\sim 5.3$. As shown in Figure
\ref{phasediagrams} for the case of nine electrons, each of these
states corresponds to a given set of momenta $\{{\bf k}_i \}$ in the Rezayi-Read
state. The Fermi seas of momenta develop in a very natural and
systematic way. Starting from an elongated sea, which we identified
as the exact solution, a single momentum is moved at each
level-crossing, terminating in a symmetric Fermi sea when
$L_1\sim L_2$.\\

Since our Luttinger liquid solution corresponds to one of the Fermi
seas in the Rezayi-Read state and this state develops smoothly
towards the bulk, we conclude that the Luttinger liquid of neutral dipoles is continuously connected 
to the bulk ground state. 

\section{Conclusions}\label{con}

We conclude that the thin torus provides a simple and accurate picture of both abelian 
and non-abelian quantum
Hall states, and even more surprisingly, also of the gapless state at
$\nu=1/2$. The gapless state is particularly important 
since it provides an explicit microscopic example of how weakly interacting quasiparticles 
moving in a reduced (zero) magnetic field emerge as the low energy sector of strongly 
interacting fermions in a strong magnetic field.

There are strong reasons to believe that the picture presented
here is valid also for other quantum Hall states. Indeed, the ground state 
and quasihole degeneracies of other topological states can be obtained 
on the thin torus \cite{read,marchmeeting}.

A one-dimensional picture of the quantum Hall system is very natural, 
and in some sense almost obvious. After all, a single Landau level is a 
one-dimensional system. The non-trivial result is, of course, that a model 
with an interaction that is short range in the one-dimensional sense is relevant. We believe that 
the evidence reviewed here establishes that this is indeed the case.

\begin{acknowledgement}
We thank Hans Hansson, Janik Kailasvuori, Emma Wikberg and Maria
Hermanns for interesting discussions and fruitful collaborations.
This work was supported by the Swedish Research Council and by
NordForsk.
\end{acknowledgement}

\end{document}